\documentclass[aps,prb,superscriptaddress,12pt,tightenlines,nofootinbib]{revtex4}

\begin{document}

\def\>{\rangle}
\def\<{\langle}

\title{An Introduction to QBism \\ with an Application to the Locality of Quantum Mechanics}

\author{Christopher A. Fuchs}
\affiliation{Stellenbosch Institute for Advanced Study, 19 Jonkershoek Road, Stellenbosch 7600, South Africa}\affiliation{Raytheon BBN Technologies, 10 Moulton Street, Cambridge MA 02138, USA}
\author{N.~David Mermin}
\affiliation{Stellenbosch Institute for Advanced Study, 19 Jonkershoek Road, Stellenbosch 7600, South Africa}\affiliation{Laboratory of Atomic and Solid State Physics, Cornell University
  Ithaca NY 14853, USA}
\author{R\"udiger Schack}\affiliation{Stellenbosch Institute for Advanced Study, 19 Jonkershoek Road, Stellenbosch 7600, South Africa}\affiliation{Department of Mathematics, Royal Holloway University of
London, Egham, Surrey TW20$\;$0EX, UK}

\begin{abstract}
We give an introduction to the QBist interpretation of quantum mechanics.  We
note that it removes the paradoxes, conundra, and pseudo-problems that have
plagued quantum foundations for the past nine decades.  As an example, we show
in detail how it eliminates ``quantum nonlocality''.
\end{abstract}

\maketitle

\section{QBism}

\begin{quote}
\noindent {\sl In our description of nature the purpose is not
 to disclose the real essence of the phenomena
 but only to track down, so far as it is possible, relations
between the manifold aspects of our experience.}

\hskip 10 pt --- Niels Bohr, 1929\cite{Bohr61}

\vskip 10pt
\noindent {\sl Physics is to be regarded not so much
as the study of something a priori given,
but as the development of methods for
ordering and surveying human experience.}

\hskip 10 pt --- Niels Bohr, 1961\cite{Bohr87}
\end{quote}


Much of what Bohr had to say about the nature of quantum physics evolved over a
thirty year span containing the 1935 paper of Einstein, Podolsky, and Rosen
[EPR].\cite{Einstein35}   But Bohr's view on the central role in science of human experience
survived the trauma of EPR more or less intact.

Although it differs in many important ways from what has come to be called ``the Copenhagen interpretation", QBism\cite{Fuchs10} --- Quantum Bayesianism  ---  agrees with Bohr
that the primitive concept of  {\it experience} is fundamental to an understanding of science.   According to
QBism, quantum mechanics is a tool anyone can use to evaluate, on the basis of one's past experience, one's probabilistic expectations for one's subsequent  experience.

Unlike Copenhagen, QBism explicitly takes the ``subjective" or  ``judgmental"
or ``personalist" view of probability,\cite{Savage54,deFinetti90,Bernardo94,Jeffrey04,Lindley06}
which, though common among contemporary  statisticians and eco\-nom\-ists, is still rare among physicists:  probabilities are assigned to an event by an agent\footnote{``Agent" in the sense of one who acts (and not in the sense of one who represents another).  We follow the widespread practice in the quantum-information community of calling the agent Alice, and a second agent she might have dealings with, Bob.} and are particular to that agent.   The agent's probability assignments express her own personal degrees of belief about the event.    The personal character of probability  includes cases in which the agent is certain about the event:  even probabilities 0 and 1 are measures of an agent's (very strongly held) belief.

The subjective view returns probability theory to its historic origins in gambling.   An agent's probabilities are {\it defined} by her  willingness to
place or accept any bets she believes to be favorable to her on the basis of those probabilities.
It is a striking, and, for most physicists, surprising fact that all of the
usual probability rules can be derived from just one requirement, known as
Dutch-book coherence: an agent's probability assignments must never place her
in a position where she necessarily suffers a loss.\cite{deFinetti90,Fuchs13}

These rules constrain the set of probabilities used by any single agent.   It makes no sense to impose Dutch-book coherence on a combination of probability assignments made by more than one agent.   In this  sense probability is a ``single-user theory'':     probability assignments express the beliefs of the agent who makes them, and refer to that same agent's expectations for her subsequent experiences.   The term ``single-user" does not, however,  mean that  different users cannot each assign their own coherent probabilities.

A measurement in QBism is more than a procedure in a laboratory.  It is {\it any} action an agent takes to elicit a set of possible experiences.   The measurement outcome is the particular experience of that agent elicited in this way. Given a measurement outcome, the quantum formalism guides the agent in updating her
probabilities for subsequent measurements.   QBism addresses John Bell's
complaint
that physics should not be limited to  the outcomes of ``piddling'' laboratory tests,\cite{Bell90} by allowing each of us to take the scope of physics to be {\it any\/} of the manifold aspects of our own experience.

A measurement does not, as the term unfortunately
suggests, reveal a pre-existing  state of affairs.   It is an action on the world by an agent that results in the {\it creation\/} of an outcome --- a new experience for that agent.  ``Intervention" might  be a better term,\cite{Fuchs10b} but after more than 80 years of ``measurement" the word is hard to purge from quantum theory.
Quantum states determine probabilities through the Born rule.     Since probabilities are the personal judgments of an agent, it follows that a quantum state assignment is also a personal judgment of the agent assigning that state.\cite{Caves02}  The notorious ``collapse of the wave-function" is nothing but the updating of an agent's state assignment on the basis of her experience.

Acting as an agent, Alice can use the formalism of quantum mechanics to model any physical system external to herself.  QBism directs her to treat {\it all\/} such external systems on the same footing, whether they be atoms, enormous molecules, macroscopic crystals, beam splitters, Stern-Gerlach magnets,  or even agents other than Alice.   In this respect, QBism differs importantly from the Copenhagen interpretation as expounded by Landau and Lifshitz.  According to their quantum mechanics text\cite{Landau65}
``It is in principle impossible$\ldots$ to formulate the basic concepts
of quantum mechanics without using classical mechanics,"
and
``By measurement, in quantum mechanics, we understand any
process of interaction between classical and quantum objects
occurring apart from and independently of any observer.''
Landau and Lifshitz regard measurement and preparation devices as belonging to a separate classical domain beyond the scope of quantum mechanics. Presumably they would treat other agents as classical too, were they not explicitly excluded from consideration by the last clause in the second quotation.

In QBism the only phenomenon accessible to Alice which she does not model with quantum mechanics is her own direct internal awareness of her own private experience.  This (and only this) plays the role of the ``classical objects" of Landau and Lifshitz for Alice (and only for Alice).   Her awareness of her past  experience forms the basis for the beliefs on which her state assignments rest.     And her probability assignments express her expectations for her future experience.

The personal  internal awareness of agents other than Alice of their own private experience is, by its very nature, inaccessible to Alice, and therefore not something she can apply quantum mechanics to.  But verbal or written reports to Alice by other agents that attempt to represent their private experiences are indeed part of Alice's external world, and therefore suitable for her applications of quantum mechanics.   Having always stressed the crucial importance of stating the results of experiments in ordinary language,\cite{Bohr54} Bohr would probably have been comfortable with Alice's indirect access to Bob's experience, through language.

But Bohr would not have approved of Alice superposing reports from Bob about his own experience, as QBism requires her to do if she wants to subject those reports to analysis before they enter her own experience.   We believe Bohr would have viewed Bob's reports --- formulations in ordinary language --- as beyond the scope of quantum mechanics.   But because Alice can treat Bob as an external physical system, according to QBism she can assign him a quantum state that encodes her probabilities for the possible answers to any question she puts to him.   When Alice elicits an answer from Bob, she treats this as she treats any other  quantum measurement.     Bob's answer is created for Alice only when it enters her experience. A QBist  does not treat Alice's interaction with Bob any differently from, say, her interaction with a Stern-Gerlach apparatus, or with an atom entering that apparatus.

This means that reality differs from one agent to another.   This is not as strange as it may sound.     What is real for an agent rests entirely on what that agent experiences, and different agents have different experiences.   An agent-dependent reality is constrained by the fact that  different agents can communicate their experience to each other, limited only by the extent that personal experience can be expressed in ordinary language.   Bob's verbal representation of his own experience can enter Alice's, and vice-versa.   In this way a common body of reality can be constructed, limited only by the inability of language to represent the full flavor --- the ``qualia" --- of personal experience.

A QBist takes quantum mechanics to be  a personal mode of thought --- a very powerful tool that any agent can use to organize  her own experience.   That each of us can use such a tool to organize our own experience  with spectacular success is an extremely important  objective fact about the world we live in.   But quantum mechanics itself does not deal directly with the objective world; it  deals with the experiences of that objective world that belong to whatever particular agent is making use of the quantum theory.

Given the unprecedented murkiness that has enveloped quantum foundations for the past nine decades it is not surprising that an unorthodox approach is needed to dispel the fog.   But QBism is not all that radical.  It just requires one to recognize and abandon a strongly established way of thinking that served us reasonably well before we started to explore realms at the atomic scale.

To illustrate how QBism illuminates quantum foundations, we use it in Sections II--V to demonstrate that, contrary to the view of some physicists and many philosophers of science, there is no clash between quantum mechanics and special relativity.   ``Quantum nonlocality" is an artifact of inappropriate interpretations of quantum mechanics.   The discussion in these sections also brings out other important features of QBism.

Section VI expands on the strongly established way of thinking that QBism replaces.

\section{Against Nonlocality}

There is no nonlocality in quantum theory;  there are only some nonlocal {\it interpretations\/} of quantum mechanics.
The most famous is Bohm\-ian mechanics,\cite{Bohm52} whose nonlocality  inspired John Bell to show that nonlocality must be a feature of any interpretation that ``completes" quantum mechanics in the sense of Einstein, Podolsky, and Rosen.

But there are also local interpretations.  Many people have pointed out that quantum nonlocality cannot be demonstrated within the many-worlds interpretation.\cite{DeutschHayden,Tipler00,Bacciagaluppi02}   Robert Griffiths has argued that quantum mechanics is local in the consistent-histories interpretation.\cite{Griffiths11}  Many of those who subscribe to some version of the Copenhagen interpretation are skeptical about nonlocality, or reject it outright.\cite{Peres04,Englert13}
QBism is also a local interpretation, but with a rather different flavor.

Many Worlds and Consistent Histories work within quantum mechanics by extracting their {\it Weltanschauungen\/} from  formal mathematical features of the theory.   In contrast to these, and in contrast to Copenhagen, QBism makes sense of quantum mechanics by taking an unfamiliar perspective on scientific theories and the scientists who use them.
QBist quantum mechanics is local because its entire purpose is to enable any single agent to organize her own degrees of belief about the contents of her own personal experience.      No agent can move faster than light:   the space-time trajectory of any agent is necessarily time-like.  Her personal experience takes place along that trajectory.

Therefore when any agent uses quantum mechanics to calculate  ``[cor]relations  between the manifold aspects of [her] experience'',  those experiences cannot be space-like separated.    Quantum correlations, by their very nature, refer only to time-like separated events: the acquisition  of experiences by any single agent.     Quantum mechanics, in the QBist interpretation, cannot assign correlations, spooky or otherwise, to space-like separated events, since they cannot be experienced by any single agent.    Quantum mechanics is thus {\it explicitly} local in the QBist interpretation.

And that's all there is to it.

\vskip 10pt

Why, then,  do  many people wrongly claim that quantum mechanics is nonlocal?   They  do so  by denying at least one of three fundamental precepts of QBism:

(1) A measurement outcome does not preexist the measurement.   An outcome is created for the agent who takes the measurement action only when it enters the experience of  that agent.  The outcome of the measurement {\it is\/} that experience.    Experiences do not exist prior to being experienced.

(2)  An agent's assignment of probability 1 to an event  expresses that agent's personal belief that the event is certain to happen.   It does not imply the existence of an objective mechanism that brings about the event.     Even probability-1 judgments are judgments.  They are judgments in which the judging agent is supremely confident.

(3) Parameters that do not appear in the quantum theory and  correspond to nothing in the experience of any potential agent can play no role in the interpretation of quantum mechanics.

We take up these points in the next three Sections.

\pagebreak

\section{Experiences do not exist prior to being experienced}

\begin{quote}
\noindent {\sl Unperformed experiments have no results.}

\hskip 10pt --- Asher Peres\cite{Peres78}
\vskip 5pt
\noindent{\sl  This experiment has no outcome until I experience one.}

\hskip 10pt --- Agent undertaking an experiment
\end{quote}


QBism personalizes the famous dictum of Asher Peres.  The outcome of an experiment is the experience it elicits in an agent.   If an agent experiences no outcome, then for that agent there {\it is\/} no outcome.
Experiments are not  floating in the void, independent of human agency.   They are actions taken by an agent to elicit an outcome.   And an outcome does not become an outcome until it is experienced by the agent.
That experience {\it is} the outcome.

This is illuminated  by the famous ``paradox" of Wigner and his friend.    
The friend makes a measurement in a closed laboratory and experiences an outcome.  Wigner, outside the laboratory, doesn't experience an outcome. If he believes what his friend has told him about her plans in her laboratory he will assign an entangled state to her, her apparatus, and the system on which she is making her intervention.  Wigner's state superposes all  the possible reports from his friend about her own experience,  correlated with the corresponding readings of her apparatus.

The disagreement between Wigner's account and his friend's is paradoxical only if you take a measurement outcome to be an objective feature of the world, rather than the contents of an agent's experience.   The paradox vanishes with the recognition that a measurement outcome is personal to the experiencing agent.    There is an outcome in the friend's experience; there is none yet in Wigner's. Of course their accounts differ.  If Wigner goes on to ask his friend about her experience, then the disagreement is resolved the moment he receives her report, i.e.~when it enters his own experience.

This is relevant to the usual nonlocality story, in which Alice and Bob agree on a particular entangled state assignment to a pair of systems, one near Alice, the other near Bob.  Each then makes a measurement on their nearby system.    In the usual story the outcomes are implicitly assumed to come into  existence at the site of each measurement at the moment that measurement is performed.

What the usual story  overlooks is that the coming into existence of a particular measurement outcome is valid only for the agent  experiencing that outcome.   At the moment of his own measurement Bob is playing the friend to Alice's far-away Wigner, just as at the moment of her own measurement she is playing the friend to Bob's Wigner.  Although each of them experiences an outcome to their own measurement, they can experience an outcome to the measurement undertaken by the other only when they receive the other's report.  Each of them applies quantum mechanics in the only way in which it can be applied, to account for the correlations in two measurement outcomes registered in his or her own individual experience.   And as noted above, experiences of a single agent are necessarily time-like separated.  The issue of nonlocality simply does not arise.

By reifying {\it all} measurement outcomes, without  reference to the agent taking the action or to that agent's subsequent experience, the usual nonlocality  arguments implicitly embrace the Copenhagen view that measurement outcomes belong to an objective (``classical") domain that is independent of agents and/or their experience.    In QBism, however, an agent applies quantum mechanics to {\it everything} outside her internal personal experience.   There is a vestigial remnant  in QBism of  the Copenhagen classical domain, but  the vestige of this ``classical domain'' varies from one agent to another  and  is limited to that agent's directly perceived personal experience.

\section{Probability-1 assignments are judgments}

\begin{quote}
\noindent {\sl The abandonment of superstitious beliefs about the existence
of  Phlogiston, the Cosmic Ether, Absolute Space and Time$\ldots,$
or Fairies and Witches, was an essential step along the road
to scientific thinking.  Probability, too,  if regarded as
something endowed with some kind of objective existence,
is no less a misleading misconception, an illusory attempt
to exteriorize or materialize our actual probabilistic beliefs.}

\hskip 10pt--- Bruno de Finetti\cite{deFinetti90} 
\vskip 10pt

\noindent {\sl  Why, when an event appears to me as practically certain (i.e., when I evaluate its probability as close to 1) have I the right to be practically certain that it will occur?   Because when I say that an event is practically certain (when I evaluate its probability as close to 1) I do not say nor can I want to say more or less than this: that I am practically certain it will occur.}


\hskip 10pt--- Bruno de Finetti\cite{deFinetti31}
\end{quote}


When Alice assigns an event $E$ probability 1, she is announcing her willingness to
buy a ticket that pays her a dollar
if $E$ happens, for  any amount less than one dollar,
and to sell a ticket requiring her to pay a dollar
if $E$ happens, for any amount more than one dollar.\footnote{If she assigns E a probability $p$, replace ``one dollar" by ``$p$ dollars".}  It does not mean that there is an objective feature of the world
that makes $E$  happen.

That probability-1 (or probability-0) judgments are still judgments, like any other probability assignments,  may be the hardest principle of QBism for physicists to accept.    But it belongs to a tradition going back at least to David Hume's critique of induction.   Why, just because something has invariably happened in the past, can we infer that it will continue to happen in the future?       The only obvious answer --- that this principle of inference has always worked in the past --- glaringly appeals to the very law of inference it has been invoked to justify.

Inductive inference is nothing more than a broadly shared personal judgment based on habit.  The habit may even be hard-wired by evolution,  but habit it remains.    Should the sun not rise tomorrow we would all have a lot of things to rethink, if we survived the catastrophe.   But appealing to the conservation of angular momentum as the objective mechanism that guarantees future sunrises misses the point.   The laws of classical mechanics are themselves codifications of
innumerable inductive inferences extracted from centuries of human experience, and as such cannot explain the success of the inductive method.  To be sure, they are extraordinarily elegant and concise codifications, but to say that is also to justify them in human terms.    This may sound better than ``habit'', but it is just as much a subjective judgment.

The mistake in the 1935 argument of Einstein, Podolsky, and Rosen lies in their taking probability-1 assignments to indicate objective features of the world, and not just
firmly held beliefs.  Their argument uses the famous EPR reality criterion:  ``If, without in any way disturbing a system, we can predict with certainty (i.e., with probability equal to unity) the value of a physical quantity, then there exists an element of physical reality corresponding to this physical quantity.''   Without such ``elements of physical reality'',  there is no basis for their argument that if quantum mechanics gives a complete description of physical reality, then what is real in one place depends upon the process of measurement carried out somewhere else.

Bohr maintained that EPR's mistake lay in an ``essential ambiguity'' in their phrase ``without in any way disturbing''.\cite{Bohr35}   But for a QBist their error is simpler.   Their mistake was their failure to understand, as many physicists today continue not to understand, that $p=1$ probability assignments are very firm personal judgments of the assigning agent, and nothing more.

The unwarranted assumption that probability-1 judgments are necessarily backed up by objective facts-on-the-ground --- elements of physical reality --- underlies EPR's conclusion that if quantum mechanics is complete then it must be (unacceptably to them) nonlocal.    It also underlies Bell's original 1964 derivation of the Bell inequalities.\cite{Bell64}

\vskip 10pt

There is, finally, a weaker kind of nonlocality argument that has a QBist  refutation different from those given above.

\section{The contents of physical science}

In Sections II and III  we noted that in QBism quantum correlations are necessarily between time-like separated events, and therefore cannot be associated with faster-than-light influences.   This removes any tension between quantum mechanics and relativity.

 ``Quantum  nonlocality'' is, however, also invoked when one event is said to influence another  in the absence of any known connecting mechanism, whether or not the two events are space-like separated.  Space-like separation is relevant only to ensure the absence of an  {\it unknown\/} mechanism of influence, since whatever that influence might be, it would then have to violate the relativistic prohibition on faster-than-light action.  But a slower-than-light influence can also be spooky, if it cannot be accounted for by any currently known mechanism.

 There are also no spooky slower-than-light influences in QBist quantum mechanics.   As Asher Peres liked to put it, nothing is propagated as a result of a measurement, neither faster than light nor slower than a snail.  Bob's system is not changed by Alice's far away intervention in any way whatsoever.

 The  simplest such attempts to demonstrate  this kind of  ``quantum nonlocality''  involve outcomes (not necessarily space-like separated) at detectors in two separate spatial regions $A$ and $B$ (presided over by Alice and Bob).\cite{Bell2004}  The outcomes are triggered by entities --- call them ``particles" --- that come to Alice and Bob from a common source $S$.   Each detector can be operated in two different settings, $i = a \ {\rm or}\ a'$, and   $j = b\ {\rm or}\ b'$, and each when triggered can produce one of two outcomes, $x = -1 \ {\rm or}\ 1$ and $y = -1 \ {\rm or}\ 1$.   Alice and Bob agree on the joint state they assign to the particles incident on their detectors, they agree on the nature of those detectors, and therefore they both use quantum mechanics to calculate the same probabilities $p(x,y|i,j)$ for the outcomes each of them experience for each of the four possible combinations of settings.

Bell inequalities are necessary conditions for these four joint probabilities all to be  of the form
$$p(x,y|i,j) =  \<p(x,y|i,j,\lambda)\>, \eqno(1)$$
where $p(x,y|i,j,\lambda)$ is of the form
$$ p(x,y|i,j,\lambda) = p(x|i,\lambda)p(y|j,\lambda),\eqno(2)$$
and the brackets $\<\ \>$  in Eq.~(1) denote a weighted average over a parameter or set of parameters $\lambda$.   The nature of the parameters $\lambda$ is rarely subject to much critical scrutiny.\footnote{Examples of this can be found in recent papers by Pusey, Barrett, and Rudolph\cite{PBR} and Colbeck and Renner\cite{Colbeck12}, along with much of the literature these papers have prompted.}   
What is agreed is that the parameters and their weights are independent of the choice $i,j$ of Alice and Bob's settings, and that conditioning on $\lambda$ removes (``screen off$\,$") all correlations between Alice's and Bob's outcomes $x$ and $y$  for  all choices of the modes $i$ and  $j$ that Alice and Bob might select for their detectors, as indicated in Eq.~(2).

If a set of distributions violates a Bell inequality, as many sets of quantum distributions do, then there can be no such sets of parameters.  What has the nonexistence of such parameters to do with nonlocality?   For a QBist  the answer is ``nothing!''~because Eqs.~(1) and (2) are talking about nothing.
The parameter $\lambda$ is undefined.  It does not appear in the quantum theory.  Nor has anybody ever suggested what in the experience of an agent $\lambda$ might correspond to. In QBism this puts it outside the scope of physical science.

What the parameter $\lambda$ expresses is a classical intuition that correlations in the experiences of agents in widely separated regions ought to find their explanation in correlations in conditions prevailing in those regions.    In particular when the local experiences are mediated by the arrival of particles originating at a common source, $\lambda$ is supposed to represent common objective features of those particles imposed on them at that source.   These features affect the outcomes Alice and Bob experience.
It is an important fact, surprising to one's classical intuition, that the correlations in Alice's and Bob's outcomes cannot be accounted for in this way.  But this does not mean that anything in Alice's experience is influenced by Bob's choice of setting, or vice-versa.    The variable $\lambda$ is nothing more than a version of the discredited EPR elements of reality. For a QBist the nonexistence of such objective facts-on-the-ground as $\lambda$ no more implies nonlocality than does the nonexistence of elements of reality in the original EPR argument.

\section{Conclusion}

\begin{quote}
\noindent {\sl One can only help oneself through something like the following emergency decree:  Quantum mechanics forbids statements about what really exists -- statements about the object.
Its statements deal only with the object-subject relation. Although this holds, after all, for any description of nature, it evidently holds in a much more radical and far reaching sense in quantum mechanics.}

\hskip 10pt --- Erwin Schr\"odinger, 1931 letter to Arnold Sommerfeld\cite{Schroedinger1931}

\vskip  12pt

\noindent {\sl  [Some]  factors crucial to the perception of scientific work [are]$\,\ldots$ that the task of science is described in full if we limit it to showing how, because of our unique organization, the world must inevitably appear to us; that the eventual results of science, precisely because of the manner of their acquisition, are conditioned not only by our organization but also by what influenced that organization; and lastly that the problem of a world constitution that takes no account of the
mental apparatus by which we perceive it is an empty abstraction,
of no practical interest.}

\hskip 10pt    --- Sigmund Freud,   {\it The Future of an Illusion}, 1927\cite{Freud1927}

\end{quote}

Bohr was not the only one to articulate elements of a QBist position in the first third of the 20th century.  Two great Viennese investigators took such a  view of science.  {\it The Future of an Illusion}   is about the origin of religious belief, but Freud concludes his essay with a thoroughly QBist characterization of our scientific understanding of the world.   Schr\"odinger explicitly takes a QBist view of quantum mechanics in his 1931 letter to Sommerfeld.   Three decades later,  in {\it Nature and the Greeks}\cite{Schroedinger1951}, he takes a  QBist view of science more generally and hardly even mentions quantum mechanics.   He stresses that because everything any of us knows about the world is constructed out of his or her individual private experience,  it can be unwise to rely on a picture of the physical world from which personal experience has been explicitly excluded, as it has been from  physical science.  Schr\"odinger traces this exclusion back more than two thousand years to the ancient Greeks.  It worked for over two millennia and played an important role in the construction of classical science.

But when we attempted to understand phenomena at scales not directly accessible to our senses, our ingrained practice of divorcing the objects of our investigations from the subjective experiences they induce in us got us into trouble.  While our efforts at dealing with phenomena at these new scales were spectacularly successful, we have just as spectacularly failed for almost a century to reach any agreement about the nature or meaning of that success.

The Founders of quantum mechanics were already aware that there was a  problem.   Bohr and Heisenberg  dealt with it by emphasizing the inseparability of the phenomena  from the instruments we devised to investigate them.   Instruments are the Copenhagen surrogate for experience.   Being objective and independent of the agent using them, instruments miss the central point of QBism,  giving rise to the notorious measurement problem, which has vexed physicists to this day.

Bohr seems closer to QBism in the two quotations at the head of Section I.  But we suspect that, if pressed to explain ``experience'', he would have fallen back on our experience of a classical apparatus. Unlike                                
 a QBist he would not give the term enough scope --- any and all experience --- or the appropriately personalistic primacy:  any user's  own experience constitutes all of the raw material out of which she constructs her world.

The subsequent view of Heisenberg\cite{Heisenberg2000} and Rudolph Peierls\cite{Peierls1991}, that quantum physics was not about the world, but about our ``knowledge" of the world, gets even closer to the real issue.  But it raises tough questions.    Whose knowledge?    Knowledge about what?   The trouble is that ``knowledge'' is the wrong term,  for two reasons.

First, it is wrong because  generally there are many different agents.   Anybody using quantum mechanics to organize her experience can be an agent, and different agents have different experiences.   ``Knowledge''   can suggest an agent-independent factuality.  ``Belief$\,$", which unavoidably  implies a believer,  is more balanced between subject and object.

Second, ``knowledge" is the wrong word because the fundamental output of the quantum theory is not a set of facts, but a set of probabilities.   The probabilities in quantum mechanics, like all probabilities,  express the willingness of the agent using them  to take or place bets.   That willingness is based on  personal judgment, informed by  the agent's beliefs.

We have noted that the QBist position, that quantum states are personal judgments of an agent, is an inevitable consequence of the subjective view of probability expressed so eloquently by Bruno de Finetti.  
On the other hand, given its ability to remove the long-standing paradoxes and absurdities that have plagued  quantum foundations for almost a century, one could invert the argument and maintain that QBism  provides a powerful validation of the personalist view of probability.    The troubled history of quantum foundations  is a telling indictment of the  proposition that probabilities, even including probabilities 1 and 0, are not personal judgments but facts, backed up by objective features of the world.

Among those absurdities is ``quantum  nonlocality".  We have been told that QBism is ``too great a price to pay" to restore locality to physics.     But a prejudice against nonlocality is not the reason for embracing QBism.   The recognition that science has a subject as well as an object liberates us from the grip of an ancient Greek maneuver that worked for over two millennia, until it tripped us up in the last  century.   Restoring the subject-object balance  clears up the obscurities and ambiguities of the Copenhagen interpretation, eliminates the measurement problem, and --- incidentally --- invalidates the claim that quantum mechanics is nonlocal, or in conflict, or just in a state of tension, with special relativity.

We bring QBism to the reader's attention because it corrects a profound misconception  in our  general view of science, which led us into major confusion in the 20th century.   Now that we are well into the 21st  and we all agree that quantum mechanics works spectacularly well for every practical purpose,  surely it is  time to expand our ancient view of the nature of science, to dispel the murkiness that has obscured the foundations of the theory for too long.

\end{document}